\documentclass{aastex631}

\newcommand{\speed}[1]{#1 km~s${}^{-1}$}

\begin{document}

\title{Ubiquitous Small-scale EUV Upflow-Like Events above Network Regions Observed by the Solar Orbiter/Extreme Ultraviolet Imager}

\correspondingauthor{Hechao Chen}
\email{hechao.chen@ynu.edu.cn}
\correspondingauthor{Yadan Duan}
\email{duanyadan@pku.edu.cn}

\author[0000-0001-9491-699X]{Yadan Duan}
\affiliation{School of Earth and Space Sciences, Peking University, Beijing, 100871, People's Republic of China}
\author[0000-0001-7866-4358]{Hechao Chen}
\affil{School of Physics and Astronomy, Yunnan University, Kunming 650500, People’s Republic of China}
\author[0000-0003-4804-5673]{Zhenyong Hou}
\affil{School of Earth and Space Sciences, Peking University, Beijing, 100871, People's Republic of China}
\author[0000-0001-5657-7587]{Zheng Sun}
\affil{School of Earth and Space Sciences, Peking University, Beijing, 100871, People's Republic of China}
\author[0000-0001-9493-4418]{Yuandeng Shen}
\affiliation{Shenzhen Key Laboratory of Numerical Prediction for Space Storm, School of Aerospace, Harbin Institute of Technology, Shenzhen 518055, People’s Republic of China}

\begin{abstract}
Universal  small-scale solar activity in quiet region are suggested to be a potential source of solar wind and the upper solar atmosphere. Here, with the high-resoltion 174 \AA~imaging observations from the Solar Orbiter/Extreme Ultraviolet Imager (EUI), we investigate 59 EUV upflow-like events observed in the quiet Sun. Their average apparent (plane-of-sky) velocity, lifetime, and propagation distance are measured as 62 $\speed$, 68.6 s and 3.94 Mm, respectively. These upflow-like events exhibit dynamic characteristics but lack base brightening, featuring a hot front and subsequent cold plasma ejection. 39\% of the EUV upflow-like events exhibit recurrent characteristics. Unprecedented high-resolution 174 \AA~observations reveal that some EUV upflow-like events exhibit blob-like fine structures and multi-strand evolutionary features, and some upflow-like events can cause localized haze-like plasma heating ahead of their spire region during the ejection process. A subset of the EUV upflow-like events covered by the Solar Dynamics Observatory reveals that they appear at the chromospheric networks. Through emission measure analysis, we found that these upflow-like events eject hot plasma of transient region or coronal temperature (an average of $\sim$10$^{5.5}$K). We suggest that EUV upflow-like events may be EUV counterparts of chromospheric spicules and/or transition region network jets, and play a role in heating localized corona above the network regions.

\end{abstract}

\keywords{Quiet sun (1322), Solar ultraviolet emission (1533), Solar corona (1483), Solar spicules (1525)}

\section{Introduction} \label{sec:intro}
The solar atmosphere is characterized by a multitude of upward and downward plasma flows, which are important in regulating the circulation of material and the energy balance within the solar atmosphere. Downflows are characterized by the return flow of plasma to the solar atmosphere, exemplified by phenomena such as sunspot supersonic downflows \citep[e.g.,][]{2018ApJ...859..158S,2022A&A...659A.107C}, coronal rain \citep[e.g.,][]{2018ApJ...864L...4L,2024ApJ...973...57Q}, supra-arcade downflows \citep[e.g.,][]{1999ApJ...519L..93M,2023MNRAS.522.4468T} and reconnection outflows \citep[e.g.,][]{2024ApJ...974..205S}. Upflows are hypothesized to be connected to the formation of the solar wind or to facilitate the transport of mass and energy within the solar atmosphere, potentially playing a key role in coronal heating \citep{2021SoPh..296...47T}. Upflows in quiet Sun regions exhibit a wide range of scales, encompassing coronal jets \citep[e.g.,][]{2016ApJ...832L...7P}, surges \citep[e.g.,][]{2024A&A...686A.218N}, spicule\citep[e.g.,][]{2017ApJ...845L..18D}, network jets \citep[e.g.,][]{2014Sci...346A.315T} at transition region temperatures, micro-jets\citep[e.g.,][]{2021ApJ...918L..20H}, jetlets \citep[e.g.,][]{2014ApJ...787..118R,2018ApJ...868L..27P,2019ApJ...887L...8P}. Uplfows/Outflows can occur at the edges of ARs, termed as AR flows. In quiet Sun regions, upflows are preferentially occur at the boundaries of network channels. 

The blue shifts observed in the spectral lines are commonly regarded as outward-propagating disturbances or upflows. Using the SUMER on board SOHO, \citet{1999A&A...346..285D} reported quiet Sun areas exhibit an average blue shift of $\sim$ 2 $\speed$ in the upper-transition-region Ne VIII 770 \AA~line. Doppler maps of the Ne VIII 770 Å spectral line, derived from raster scans, reveal pronounced blue shifts at the junctions of multiple adjacent network structures, reaching speeds of up to 5-10 km/s \citep{1999Sci...283..810H,1999A&A...346..285D,2008ApJ...674.1144A,2008A&A...478..915T}.These upflows are considered to be signatures of solar wind outflows \citep{1999Sci...283..810H} or mass supply to quiet coronal loops \citep{2007A&A...468..307H,2008A&A...478..915T}. Using five spectral lines from SUMER (Si IV 1402 \AA~, C IV 1548 \AA~, N V 1238 \AA~, O V 1031 \AA~, and Ne VIII 770 \AA~), \citet{2009ApJ...707..524M} suggested that strong upflows, with speeds ranging from 40 to 100 km/s, appear to occur in all over the quiet Sun in and around network regions. 

\par
These upflows are spatiotemporally correlated with and have similar upward velocities as chromosphere (type II) spicules or rapid blueshifted events \citep{2009ApJ...701L...1D,2009ApJ...705..272R}. Based on the spectral characteristics, rapid blueshifted events are considered to be the solar counterparts of ondisk spicules \citep{2008ApJ...679L.167L,2009ApJ...705..272R}. Spicules are a significant component of the chromosphere, characterized by widths typically ranging from 0.3 to 1.5 Mm. They possess temperatures approximately 5 to 15$\times$10 $^3$ K, and feature a density of about 2.2$\times$10 $^{10}$ $cm^{-3}$ \citep[e.g.,][]{2020ApJ...888L..28S}. The lifetimes of type I spicules are roughly between 3 to 10 minutes, with maximum heights between 6.5 and 9.5 Mm, and their velocities generally within the range of 15 to 40 $\speed$; Type-II spicules in the chromosphere have considerably shorter lifetimes, lasting only a few tens of seconds, but they can achieve significantly higher speeds, reaching up to 150 $\speed$ \citep[e.g.,][]{1968SoPh....3..367B,2007PASJ...59S.655D,2012ApJ...759...18P,2014ApJ...792L..15P,2021A&A...647A.147B}. Type I spicules are mostly driven by p-mode wave leakage. Along magnetic flux concentrations, photospheric oscillations and convective motions can leak into the chromosphere, where they form shock waves that propel spicules upward \citep{2004Natur.430..536D,2007PASJ...59S.655D}. According to \citet{2017Sci...356.1269M},   the violently released magnetic tension can cause type II spicules, which is amplified and transported upward through ion–neutral interactions or ambipolar diffusion. Additionally, magnetic reconnection is also responsible for the production of type II spicules \citep{2007PASJ...59S.655D,2019Sci...366..890S}.
\par 
\citet{2014Sci...346A.315T} found that Type II spicules manifest as fast, intermittent network jets in transition region images. This viewpoint is further supported by \citet{2014ApJ...792L..15P,2015ApJ...799L...3R}, who argue that network jets are the on-disk counterparts and heating signatures of type II spicules. These network jets emerge from the transition region network at the base of corona, which can be observed in coronal hole \citep[e.g.,][]{2014Sci...346A.315T,2022A&A...660A.116G,2019SoPh..294...92Q} or quiet region \citep[e.g.,][]{2016SoPh..291.1129N,2018A&A...616A..99K}. In the quiet region, the network jets have an apparent speed of approximately 100$-$180 $\speed$ with lifetimes of 55$-$155 s and lengths of 2 $-$4.2 Mm \citep{2018A&A...616A..99K}. Magnetic reconnection is considered a plausible mechanism for the formation of these network jets \citep[e.g.,][]{2019ApJ...873...79C}. Based on a 2D MHD simulation, \citet{2018ApJ...852...16Y} showed that magnetic reconnection between a magnetic loop and a background open flux in the transition region can simultaneously launch a fast, warm jet similar to network jets and a slow, cool jet similar to classical spicules.

\par
The launch of Solar Orbiter (SO; \citealt{2020A&A...642A...1M}) provides us with an unprecedented opportunity to investigate small-scale solar activity in the quiet solar corona. For example, initial observations from the Extreme Ultraviolet Imager (EUI; \ \citealt{2020A&A...642A...8R}) aboard Solar Orbiter have unveiled widespread small-scale EUV brightenings in the quiet Sun, termed "campfires" \citep{2021A&A...656L...4B}, which can appear in the form of loops or individual points of brightens \citep[e.g.,][]{2021A&A...656L...4B,2021A&A...656L...7C,2021ApJ...921L..20P,2022A&A...660A.143K}. Additionally, small-scale jets or microjets observed in quiet Sun regions using the Solar Orbiter's Extreme Ultraviolet Imager exhibit projected speeds, widths, and lengths in the approximate ranges of 20-170 km/s, 0.3-3.0 Mm, and 2-20 Mm, respectively \citep{2021ApJ...918L..20H,2024A&A...686A.279S}. These small-scale jets exhibit characteristics such as footpoint brightening or an inverted Y-shaped morphology. 

 \par 
In this article, we focus on another class of outward-propagating disturbances in quiet Sun region. Diffierent from the above-mentioned microjets and campfires, these EUV upflow-like events frequently occur in quiet regions without obvious footpoint or base brightening, but they also appear to be important due to their ubiquity. We conducted a statistical investigation of 59 small-scale EUV upflow-like events in the quiet Sun, utilizing observations from the Extreme Ultraviolet Imager provided by the Solar Orbiter. They display surge-like activity, characterized by a hot front followed by the ejection of cold plasma. In previous studies, \citet{2023A&A...674A.219S} reported two upflow events in quiet regions, attempting to determine the causes and associated activities of the weak upflows observed in the quiet Sun. This paper primarily focuses on the source region magnetic field, dynamics, fine structures, and radiative characteristics of these upflows. Our results would further continue to fuel ongoing exploration about the physicial relationship between these upflow-like events and other typical or traditional solar jetting phenonema in the future.

\section{Observations} \label{sec:instr}
These 59 upflow-like events are derived from two datasets, observed in the 174 \AA~and 1216 \AA~(Ly$\alpha$) wavelengths by the EUI. The 174 \AA~band primarily focuses on the coronal line Fe X (with a core temperature of approximately 1 MK), and the Fe X line is wide enough to include other transition region lines \citep{2024arXiv240410577P}. 1216 \AA~is dominated by the Lyman-$\alpha$ line of hydrogen \citep{2021A&A...656L...4B}. One of the two datasets is for 2021 February 23, with a duration of $\sim$ 5.5 minutes, a time resolution of 2 seconds, and a pixel resolution of 0\arcsec.49. The other dataset is from 2020 May 30, with a duration of $\sim$ 4.5 minutes, a time resolution of 5 seconds, and a pixel resolution of 0\arcsec.29. In these two observations, Solar Orbiter was located at distance of about 0.52 and 0.61 astronomical units from the Sun, respectively.
\par
It should be noted that the field of view of the Solar Dynamics Observatory (SDO; \citealt{2012SoPh..275....3P}) effectively covered the data of 2020 May 30. We also used 171, 304, 131, 193, 211, and 1600 \AA~ images from the Atmospheric Imaging Assembly (AIA; \citealt{2012SoPh..275...17L}) on board the SDO to study the small-scale upflow-like events. The SDO/AIA EUV and UV images utilized in this study possess a pixel resolution of 0\arcsec.6 and a cadence of 12 s (24 s). The Helioseismic and Magnetic Imager (HMI; \citealt{2012SoPh..275..207S})  on board the SDO was provided photospheric magnetograms evolution, with a pixel resolution of 0\arcsec.5 and a cadence of 12 s. On 2020 May 30, the angle between the Solar Orbiter/EUI and SDO/AIA was approximately 10 degrees. We used cross-correlation for data alignment and jitter elimination.
\section{RESULTS} \label{sec:OBS}

\subsection{Overview of Two Datasets}
Figure 1 (a) and (b) show the two datasets as imaged by the 174 \AA~channel of Solar Orbiter. Overview imagery reveals that these upflow-like events originate from the solar quiet region. Within the 32 white boxes that we have marked, a total of 59 upflow-like events can be observed, see Table 1. We carefully checked all the regions and found that these small-scale upflow-like events originate from boundaries of network magnetic patches in the photosphere (as described in Section 3.3)\citep[e.g.,][]{2014Sci...346A.315T,2016SoPh..291.1129N,2018A&A...616A..99K,2019ApJ...873...79C,2019SoPh..294...92Q,2022A&A...660A.116G} and are common in solar quiet region. Due to the scarcity of spectral observations and considering that these upflows occur on very small spatial scales and exhibit outward-propagating disturbances behavior along magnetic field lines, although they do not always show significant brightness variations, and given that these disturbances are observed by Solar Orbiter in EUV wavelength, we classify the observed results as “EUV upflow-like events” to differentiate these features from microjets (or small-scale jets) and conventionally defined upflows observed by spectroscopy.

\par
Figure 1 (c)$-$(h) display eight EUV upflow-like events, which are marked by black arrows. It is possible that multiple EUV upflow-like events may occur within each box, as shown in animation1 (c)$-$(h). For comparison, the white arrow in panel (h) shows a brightening akin to what was shown due to the braiding of loops in \citet{2022A&A...667A.166C}. However, the EUV upflow-like events we focus on here are collimated plasma ejections, with a majority of these ejections carrying cooler material as denoted by black arrows in Figure 1 (c)$-$(h) and the animation1.

\subsection{Dynamic Characteristics, Parameters and Fine structure of the EUV upflow-like events}
To derive the projected speed, propagation duration, and propagation distance, we plotted time-distance images along the direction of plasma ejections for 59 EUV upflow-like events. Figure 2 shows four selected examples, two of which are non-repeated EUV upflows, as displayed in Figure 2 (a)$-$(b). While another two from Figure 2 (c)$-$(d) are repeated ones. In this case, we observed the continuously brightened material propagating in the direction of the arrows in Figure 2 (c)$-$(d). The time interval is about 52-55 s in these two events, and the average speed of continuous ejection is 94$\speed$ (c) and 74 $\speed$ (d), respectively. Out of the 59 examples, 61\% were classified as non-repeated EUV upflow-like events, while 39\% were classified as repeated EUV upflow-like events. This result is shown in the histogram in Figure 3 (e). Although the proportion of these repeated EUV upflow-like events is small, they exhibit the same recurrent nature as the chromospheric spicules \citep[e.g.,][]{2004Natur.430..536D}, macorospicules \citep[e.g.,][]{2023ApJ...942L..22D}, surge-like activities \citep[e.g.,][]{2018ApJ...854...92T} and coronal jets \citep[e.g.,][]{2015ApJ...815...71C,2016ApJ...833..150L}.

\par
Figure 3 also shows the other physical parameters, such as duration, projected speed, ejection distance (length), and the relationship between ejection distance and projected speed. For the non-repeated upflow-like events, the duration (distance) is calculated as the difference in time (length) between the start and end times (points) of the EUV upflow-like event trajectory marked by dashed red lines in the time-distance images, as shown in Figure 2 (a-2)$-$(b-2). The velocity of the EUV upflow-like events can be obtained by linearly fitting the bright stripes on the time-distance plot. 
For the repeated upflow-like events, the calculated duration, velocity, and ejection distance are obtained by taking the average values from multiple ejections. The averaged duration, projected velocity, and propagation distance are 68.6 s, 62 km/s, and 3.94 Mm, respectively, as displayed in Figure 3 (a)$-$(c). Figure 3 (d) demonstrates that the larger the ejection velocity of the EUV upflow-like events, the farther they propagate. From the statistical results, repeated and non-repeated EUV upflow-like events observed by EUI exhibit a very similar velocity distribution, as shown in Figure 3 (e). In addition, the intriguing distribution of repetition time gaps for the repeated EUV upflow-like events was also calculated, and the results are shown in Figure 3 (f). The averaged repeated time gap between each EUV upflow-like events is $\sim$ 84 s.
\par
Figure 4 displays three representative examples, showing the fine structure and evolution of these EUV upflow-like events. 
When the EUV upflow-like event is accompanied by the ejection upward of dark material, it can be observed that there is some brightening plasma flow on its flank, as denoted by the white arrows in Figure 4 (a). Then, what is very interesting is that the brightening plasma flows within this EUV upflow-like event collectively result in a haze-like plasma heating in a small spatial region, which can be seen at 17:13 UT in Figure 4 (a) and animation 1 (c). This is similar to the coronal counterparts, as shown by \citet{2019Sci...366..890S}. In their supplementary fig.S10, it can be seen that in the composite image of H$\alpha$ wind and AIA 171\AA~ band, the coronal counterpart appears at the front of the chromospheric spicules, showing brightening characteristics. That means the tops of these chromospheric spicules may give rise to an enhanced emission in Fe IX/X channel. In addition, we found that some blob-like (or dot-like) features of fine structure can be seen during the ejection of these EUV upflow-like events, as shown in Figure 4 (b)$-$(c). These results indicate that the fine dynamics of these EUV upflow-like events, which may be rarely observed, may not be just simple ejection under EUI high-resolution observation \citep[e.g.,][]{2021ApJ...918L..20H,2024A&A...686A.279S}.

\subsection{Source Region Magnetic Fields of the EUV upflow-like events}
Figure 5 shows five examples from 29 EUV upflow-like events in dataset 2 that were jointly observed by EUI and SDO. These upflow-like events observed by EUI also exhibit enhanced radiation response in the Ly$\alpha$ line, as displayed in Figure 5 (a-1)$-$(e-1), suggesting that these upflow-like events might be generated at lower height \citep[e.g.,][]{2021ApJ...918L..20H}. Figure 5 (a-2)$-$(e2) also displays the five examples in AIA 171\AA~,1600\AA~images, HMI magnetograms. By combining the observations from EUI and SDO/AIA images, it is evident that these upflow-like events appear at or in close proximity to network lanes characterized by enhanced emission in Ly$\alpha$ images.They are surrounded by magnetic flux concentrations with a dominant polarity, which can also be seen in animation 2. Furthermore, we examined the 29 events in dataset 2 and found that all the upflow-like events occur in network fields surrounded by magnetic flux concentrations with a dominant polarity. Considering the similar origin locations and Ly$\alpha$ radiation features observed in dataset 1, we collectively refer to all the upflow-like events observed by EUI at or in close proximity to network fields as EUV upflow-like events. 
\par
The change in magnetic fields in the source region of the five EUV upflow-like events is depicted in the sixth column of Figure 5. We found that 76\% (22 of 29) EUV upflow-like events originate from the bipolar region and only 24\%(7 of 29) upflow-like events from unipolar, see Table 1. This implies that the majority of the EUV upflow-like events start in mixed-polarity regions and have plausible signatures of flux cancellation or emergence.

\subsection{Radiation Characteristics of the Bright Front of EUV upflow-like events}
To provide a more persistent observation and investigate the ejection trajectories of the EUV upflow-like events, we made time-space diagrams by using AIA 304, 171, 131, 193, and 211 \AA~ images, such as the three examples in Figure 6. As observed in AIA 304 and 171 \AA~images, it can be seen that a majority of EUV upflow-like events consist of a bright front and a following dark plasma ejection. The bright front appears at the very beginning of EUV upflow-like events. We examined all the data available in SDO corresponding to EUI observations and found that the falling motion of the bright front and dark body can be detected in most EUV upflow-like events. The majority exhibits similar responses in the AIA 131, 193, and 211 \AA~channels, as bright fronts and subsequent dark plasmas can be observed, while there is no signal in the AIA 94 and 335 \AA~passbands. The observations of AIA also suggest that the number of repeated EUV upflow-like events should be underestimated in EUI observations. 

Assuming the bright front of the EUV upflow-like event is an isothermal structure, we can roughly estimate its temperature in Figure 7 using the EM-loci (Emission Measure) technique \citep[e.g.,][]{2011A&A...535A..46D,2012ApJ...758...54T,2014ApJ...790L..29T,2014ApJ...797...88H,2020ApJ...899...19C}. The EM-loci curves are obtained by dividing the AIA background-subtracted intensities by the temperature response function. The result suggests that the averaged temperature and the averaged emission measure (EM) of the bright front are $\sim$10$^{5.5}$K and $\sim$5.6$\times$10$^{26}$ cm$^{-5}$, respectively. This result suggests that the bright front of EUV upflow-like events has, at the very least, a transition region temperature.

\section{Discussion}
In this study, we focus on the prevalent EUV upflow-like events observed by EUI, which exhibit dynamic ejections along magnetic field lines occurring in quiet regions without significant footpoint or base brightening. Moreover, the majority of EUV upflow-like events carry cold material during the ejection process. This paper primarily investigates the magnetic fields, dynamics, fine structures, and radiative characteristics of these upflow-like events. Our research will further propel the ongoing investigation into the physical relationships between these upflow-like events and other typical or traditional solar jet-like phenomena.
\par
The speeds, lifetimes, and lengths of the upflow-like events in the corona are comparable to those of the common network jets observed in the transition region.
The averaged velocities of the EUV upflow-like events is 62 $\speed$, which is similar to the velocities of 20-70 $\speed$ obtained by \citet{2022A&A...660A.116G} for network jets in the solar quiet region. The averaged lifetime of the EUV upflow-like events is 68.6 s, which is within the range of 53.7-157.2 s obtained for quiet region network jets in the previous work of \citet{2018A&A...616A..99K}. The length of propagation distance in our work is 3.94 Mm, similar to previous studies on network jets \citep[e.g.,][]{2014Sci...346A.315T,2016SoPh..291.1129N,2018A&A...616A..99K}. The positive correlation between speed and distance is also consistent with the result of \citet{2018A&A...616A..99K}. Through the EM-loci analysis, we found that the temperature of the plasma outflows of these EUV upflow-like events is about 10$^{5.5}$ K, which is similar to the temperature of the network jets previously found using Si IV (formed temperature of $\sim$10$^5$ K ) \citep[e.g.,][]{2014Sci...346A.315T}.
\par
As suggested by \citet{2014ApJ...790L..29T}, network jets are likely the counterparts and transition region manifestations of type II spicules. Within the temperature range of 10,000 to 2 MK, \citet{2009ApJ...701L...1D} suggested the widespread presence of weak, correlated upflow-like events with velocities of order 50–100 $\speed$ across various magnetic field structures, including active regions, quiet region, and coronal holes. They considered that these upflow-like events are related to type II spicules. In our observations, the EUV upflow-like event exhibits speeds of 62 $\speed$ and a duration time of 68.6 s, which is within the range of velocities (30-110 $\speed$) and lifetimes (50-150 $\speed$) reported for Type II spicules \citep[e.g.,][]{2012ApJ...759...18P}.
The chromospheric plasma can be accelerated upwards in the form of fountain-like jets or spicules. Most of the plasma can be heated to transition region temperatures \citep[e.g.,][]{2014ApJ...792L..15P,2015ApJ...799L...3R}, while a small fraction can be heated to temperatures exceeding 1 MK, which can be observed in the EUV 171 \AA~\citep[e.g.,][]{2012ApJ...750L..25J,2019Sci...366..890S} or 211 \AA~\citep[e.g.,][]{2011Sci...331...55D} wavelength channels. Recently, \citet{2023ApJ...944..171B} also showed signature of chromospheric spicules and their associated propagating disturbances in the 193 channel as well. Additionally, our EUI observations reveal that at least 23 of the 59 observed EUV upflow-like events exhibit recurrent behavior. Further analysis incorporating SDO data suggests that the prevalence of recurrent EUV upflow-like events may be underestimated. This implies that the majority of EUV upflow-like events share characteristics similar to spicules, displaying a recurrent nature. 
\par
The EUV upflow-like events observed in our observations exhibit characteristics in the EUI 174 \AA~images similar to their coronal counterpart, as shown in the supplement fig.S10 of \citet{2019Sci...366..890S}. In their study, these coronal counterparts appearing at the tops of the chromospheric spicules, display brightening features and upward propagation followed by a backward motion.  Such bright parabolic tracks are also reminiscent of dynamic fibrils, as reported by \citet{2016ApJ...817..124S,2023A&A...670L...3M,2023A&A...678L...5M}. The observed heating fronts may be caused by the rapid dissipation of currents propagating from the photosphere through the spicules into the corona, potentially constituting a component of the spicule formation process, as simulated by \citet{2017ApJ...845L..18D}.

Using high-resolution EUI data, we found that during EUV upflow-like event ejections, brightening plasma can appear along the flanks of the EUV upflow-like events, leading to localized haze-like plasma heating. This phenomenon may be attributed to group or collective behavior associated with spicules \citep[e.g.,][]{2011Sci...331...55D,2017ApJ...849L...7D,2014ApJ...792L..15P,2023ApJ...944..171B}. These upward plasma flows could hit the surrounding medium thermalize to produce the brightening front (or the haze-like plasma heating) \citep[e.g.,][]{2019A&A...628A...8P,2020ApJ...901..148G,2021A&A...656L..13C}.

Based on the analysis of the radiation characteristics of the EUV upflow-like events brightening front, we found that their average temperature of the brightening front is approximately10$^{5.5}$K. Moreover, the EUV upflow-like events exhibit blob-like fine structures and multi-strand evolutionary features in their spire regions. These blob-like structures may arise from a compressed shock front resulting from p-mode leakage into the corona \citep[e.g.,][]{2016ApJ...817..124S,2023A&A...678L...5M}. Alternatively, they could be due to the heating fronts as hypothesized by their 2.5D numerical simulations of \citet{2017Sci...356.1269M} and \citet{2017ApJ...845L..18D}. These simulations indicate the sudden release of tension drives strong flows leading to heating of the plasma trough ion-neutral interactions or ambipolar diffusion. Furthermore, the blob-like features may be related to the magnetic reconnection, similar to the thermal evolution and fine dynamics observed of chromospheric jets (approximately 0.1 Mm in size) \citep{2012ApJ...759...33S}, coronal jets (about 3 Mm in diameter) \citep{2014A&A...567A..11Z} or coronal microjets\citep{2021ApJ...918L..20H}. According to the simulation by \citet{2013ApJ...777...16Y}, plasmoids generated during the reconnection process due to tearing instability are consistent with the bright moving blobs observed in chromospheric jets. However, our observations do not definitively confirm whether the blob-like plasma in EUV upflow-like events is caused by magnetic reconnection resulting from tearing instability.

\par
The 29 events in Dataset 2 occur exclusively within network fields surrounded by concentrated magnetic flux with dominant polarities. Similarly, in Dataset 1, the 30 events are located within or near network channels characterized by enhanced emission in the Ly$\alpha$ images, resembling the positions of microjets  \citep[e.g.,][]{2021ApJ...918L..20H} and jetlet-like events \citep[e.g,][]{2019ApJ...887L...8P}. However, it is interesting that the EUV upflow-like events showed the presence of cold material during the ejection process, which is different from the observations of microjets, jetlet-like and campfire-like events \citep[e.g.,][]{2021A&A...656L...4B,2021A&A...656L...7C,2021A&A...656L..16M,2021ApJ...921L..20P}. Obviously, the cold plasma is from the lower solar atmosphere, such as the chromosphere or transition region. Some previous works have indicated that minifilament eruptions are the fundamental driver of many coronal jets \citep[e.g.,][]{2012ApJ...745..164S,2015Natur.523..437S,2017Natur.544..452W,2017ApJ...851...67S}, in which cool material and untwisting motions from erupting minifilaments are suggested to explain the appearance of cool component and rotating motions in jet spires \citep[e.g.,][]{2011ApJ...735L..43S,2013RAA....13..253H,2021ApJ...911...33C,2021RSPSA.47700217S,2024ApJ...968..110D}. More recently, \citet{2020ApJ...893L..45S} speculated that erupting microfilaments might drive enhanced spicular activities. These microfilaments should have a smaller length than tiny minifilaments ($<$ a few Mm) as reported by \citet{2020ApJ...902....8C}. Recently, these upflow-like events have also been speculated to be associated with minifilament \citep{2023A&A...674A.219S}. In our observations, these EUV upflow-like events carry cold material that might originate from microfilaments. Because weak flux cancellation occurring in the source regions of network jets can provide a desirable brith place for miniature filaments \citep{2017ApJ...844..131P,2020ApJ...902....8C}. In addition to that, the darker material below the haze may also be due to surges. In the future, coordinated diagnostic studies of the chromosphere, transition region, and photospheric magnetic fields are needed to confirm what the darker material in EUV upflow-like eventsis coming from.

\par
Through analysis of the magnetic field source region, we found that 76\% (22 of 29) EUV upflows start in mixed polarity regions and are related to magnetic flux emergence or magnetic cancellation. In this case, the driving mechanism of EUV upflows may be driven by (repeated) magnetic reconnection as opposite-polarity magnetic elements are convected or emerge to the dominant polarity at networks, similar to the driving mechanism of type II spicules in the solar quiet region mentioned by \citet{2019Sci...366..890S}. The reconnection of network and/or plage fields with turbulent magnetic fields may create conditions that facilitate the formation of EUV upflow-like events. However, possibly due to a lower reconnection height, they do not exhibit footpoint brightening in the corona.
In addition, we also found that a few EUV upflow-like events (24\%; 7 of 29) do not accompany magnetic flux emergence or cancellation. They are found in close vicinity of unipolar plage/network regions, similar to spicules \citep[e.g.,][]{2009ApJ...705..272R,2016ApJ...820..124H,2015ApJ...802...26K,2021A&A...647A.147B}. In this scenario, the p-mode wave leakage \citep[e.g.,][]{2004Natur.430..536D}, or the release of amplified magnetic tension due to ambipolar diffusion \citep{2017Sci...356.1269M,2020ApJ...889...95M}, or torsional motions in the photosphere \citep{2017ApJ...848...38I} could also be responsible for spicule generation. In the future, combining EUI and high-resolution GST magnetic field observations to conduct observational studies is needed in the future.

\par
\section{Summary}
With high-resolution EUI 174 observation provided by Solar Orbiter, we conducted a statistical investigation on 59 small-scale EUV upflow-like events from two datasets in the quiet Sun. We explore the physical connections between these ubiquitous coronal upflow-like events in quiet regions and solar phenomena such as chromospheric spicules and transition region network jets, providing insights into their potential contribution to the heating of the localized corona above network regions. We found that the EUV upflow-like events exhibit fine dynamics in corona from EUI images, and their repeated production may heat up the local coronal above the network field. Through the analysis of the magnetic origin, dynamic process, and radiation characteristics of EUV upflow-like events, the main findings of our study are as follows: 
\begin{description}
\item[(i)]{These ubiquitous, small-scale EUV upflow-like events are consistently observed in quiet Sun regions dominated by network magnetic fields. In EUI 174 \AA~images, these upflow-like events do not exhibit significant footpoint brightening and are often characterized as small-scale dynamic features with two components: a hot front and a following cold plasma ejection.}

\item[(ii)]{The averaged velocity, duration, and propagation distance for all 59 EUV upflow-like events are 68.6 s, 62$\speed$, and 3.94 Mm, respectively.}

\item[(iii)]{At least 23 of 59 EUV upflow-like events are repeated events. The averaged time gap between repeated EUV upflows is $\sim$84 s, which is shorter than the typical 3-min or 5-min period of p-mode leakages.}
\item[(iv)]{High-resolution EUI images reveal that some EUV upflow-like events have blob-like fine structures and multi-strand evolutionary features. Some EUV upflow-like events events can cause a haze-like plasma heating in a small region at the leading edge of their ejections.}

\item[(v)]{For 29 SDO-observed EUV upflow-like events in dataset 2, most of them have obvious responses in AIA 304, 171, 131, and 211 passbands, indicating that they at least have hot plasma component at the transition-region temperature. The averaged temperature and the averaged EM of the bright front of EUV upflow-like events, derived by the EM loci method, are approximately 10$^{5.5}$K and 5.6$\times$10$^{26}$ cm$^{-5}$, respectively.}

\end{description}

\begin{acknowledgments}
The authors are grateful for the anonymous referee’s valuable suggestions and helpful discussions with Dr. Yuhang Gao from Peking University. This work is supported by National Key R\&D Program of China No. 2022YFF0503800, Beijing Natural Science Foundation (1244053),  
Postdoctoral Fellowship Program of CPSF (GZC20230097), and China Postdoctoral Science Foundation (2023M740112).  H.C.C. is supported by the NSFC grant 12103005 and the Yunnan Key Laboratory of Solar Physics and Space Science under the No. YNSPCC202210, as well as the Yunnan Provincial Basic Research Project (202401CF070165). Z.Y.H. was supported by NSFC grant 12303057. Y.D.S. was supported by the Natural Science Foundation of China (12173083). The authors are grateful for the data provided by the SDO and Solar Orbiter science teams. Solar Orbiter is a mission of international cooperation between ESA and NASA, operated by ESA. The EUI instrument was built by CSL, IAS, MPS, MSSL/UCL, PMOD/WRC, ROB, LCF/IO with funding from the Belgian Federal Science Policy Office (BELSPO/PRODEX PEA 4000112292); the Centre National d’Etudes Spatiales (CNES); the UK Space Agency (UKSA); the Bundesministerium für Wirtschaft und Energie (BMWi) through the Deutsches Zentrum für Luftund Raumfahrt (DLR); and the Swiss Space Office (SSO).
\end{acknowledgments}

\begin{deluxetable}{lccccccccccccc}
\tabletypesize{\scriptsize}
\tablewidth{0pt} 
\tablecaption{Basic properties of the EUV upflows \label{tab:deluxesplit}}
\tablehead{
\hline
\colhead{} & \colhead{Events} &\colhead{Upflows} & \colhead{Location} &  \colhead{Start time} & \colhead{Duration} & \colhead{Velocity} &\colhead{Length} & \colhead{Repeated} & \colhead{Time gap} & \colhead{Mixed-polarity}&\colhead{Network} & \\
\colhead{ } &\colhead{ } &\colhead{ }&\colhead{Pixel(X/Y)} &\colhead{(UT)} &\colhead{(s)} &\colhead{($\speed$)} &\colhead{(Mm)} &\colhead{} &\colhead{(s)} &\colhead{} &\colhead{} 
} 
\startdata 
{  dataset 1     }&E1 & 1 &    106/339&  17:06:35&76&35 &   2.67&    no &   \dots&  \dots  & \dots  \\ 
{  2021-02-23   }&E1 & 2 &    98/332&    17:04:34&52&52& 4.78&    3&    94&    \dots  & \dots  \\ 
{     }& E1& 3 &    110/243&    17:04:40&55&  87&    4.63&    4&   46&    \dots  & \dots  \\ 
{     }& E2& 4 &    1071/243&    17:04:35&56&  85&    4.8&    3&   88&    \dots  & \dots  \\ 
{     }& E3& 5 &    1740/705&    17:06:04&93&  56&    5.28&    no&  \dots&    \dots  & \dots  \\ 
{     }& E4& 6 &    693/1151&    17:05:48&74&  45&    3.31&    no&   \dots&    \dots  & \dots  \\ 
{     }&E4 & 7 &    682/1161&    17:06:00&58&  68&   4&    no&   \dots&    \dots  & \dots  \\ 
{     }& E5& 8 &    1946/1748&    17:05:18&55&  74&   3.75&   3&   90&    \dots  & \dots  \\ 
{     }& E6& 9 &    1867/893&    17:07:45&114&  44&    4.91&    no&  \dots&    \dots  & \dots  \\ 
{     }& E6 & 10 &    1857/893&    17:04:49&81&  43&    3.51&   2&  64&    \dots  & \dots  \\ 
{     }& E7 & 11 &    1408/253&    17:06:16&46&  81&    3.52&   2&  91&    \dots  & \dots  \\ 
{     }& E7& 12 &    1392/255&    17:04:55&46&  134&    6.04&   4&  68&    \dots  & \dots  \\ 
{     }& E8 & 13 &    534/1813&    17:04:40&60&  68&    4.02&   no&  \dots&    \dots  & \dots  \\ 
{     }& E8 & 14 &   563/1830&    17:05:00&51&  42&    2.06&   2&  95&    \dots  & \dots  \\ 
{     }& E8 & 15 &    536/1849&    17:05:20&36&  101&    3.7&   2&  90&    \dots  & \dots  \\ 
{     }& E9 & 16 &    1003/789&    17:04:22&90&  80&    7.24&   no&  \dots&    \dots  & \dots  \\ 
{     }& E9 & 17 &    957/802&    17:04:20&79&  62&    4.88&   no&  \dots&    \dots  & \dots  \\ 
{     }& E10 & 18 &    99/481&    17:04:52&94&  35&    3.32&   no&  \dots&    \dots  & \dots  \\ 
{     }& E10 & 19 &    126/456&    17:05:41&67&  46&    3.15&   2&  135&    \dots  & \dots  \\ 
{     }& E11 & 20 &    514/1325&    17:05:44&63&  40&    2.51&   no&  \dots&    \dots  & \dots  \\ 
{     }& E12 & 21&    1487/1743&    17:04:52&66&  56&    3.5&   3&  127&    \dots  & \dots  \\ 
{     }& E13 & 22 &    1797/1561&    17:07:10&71&  49&    3.41&   2&  61&    \dots  & \dots  \\ 
{     }& E14 & 23 &    197/933&    17:06:59&101&  36&   3.63&   no&  \dots&    \dots  & \dots  \\ 
{     }& E15 & 24 &    375/433&    17:04:59&54&  44&    2.33&   3&  79&    \dots  & \dots  \\ 
{     }& E15 & 25 &    360/396&    17:05:12&40&  89&    3.4&   no&  \dots&    \dots  & \dots  \\ 
{     }& E16 & 26 &    251/362&    17:07:29&37&  81&    2.83&   2&  51&    \dots  & \dots  \\ 
{     }& E16 & 27 &    262/352&    17:06:13&70&  42&    3.01&   no&  \dots&    \dots  & \dots  \\ 
{     }& E16 & 28 &    288/356&    17:05:54&29&  179&    3.83&   2&  121&    \dots  & \dots  \\ 
{     }& E17 & 29 &    1208/60&    17:06:56&47&  121&    5.39&   2&  123&    \dots  & \dots  \\ 
{     }& E17 & 30 &    1216/52&    17:09:05&66&  44&    2.8&   no&  \dots&    \dots  & \dots  \\ 
\hline
{  dataset 2   }& E18 & 31 &    804/777&    14:52:02&80&  46&    3.72&   no&  \dots&    no  & yes  \\ 
{  2020-05-30   }& E18 & 32 &    815/797&    14:50:43&63&  38&   2.4&   no&  \dots&    no  & yes  \\ 
{   }& E18 & 33 &    829/777&    14:52:13&84&  28&   2.37&   no&  \dots&    no  & yes  \\ 
{   }& E19 & 34 &    1130/1327&    14:51:41&57&  166&   9.78&   no&  \dots&    yes  & yes  \\ 
{   }& E19 & 35 &    1157/1354&    14:51:09&81&  130&   10.5&   no&  \dots&    yes  & yes  \\ 
{   }& E20 & 36 &    739/1017&    14:50:11&65&  55&   3.54&   no&  \dots&    no  & yes  \\ 
{   }& E20 & 37 &    723/1019&    14:50:27&64&  49&   3.04&   no&  \dots&    no  & yes  \\ 
{   }& E20 & 38 &    732/1000&    14:51:35&82&  69&   5.69&   no&  \dots&    no  & yes  \\
{   }& E21 & 39 &    680/1434&    14:50:16&81&  57&   4.58&   3&  40&    yes  & yes  \\
{   }& E21 & 40 &    678/1443&    14:50:43&81&  69&   5.6&   no&  \dots&   yes  & yes  \\
{   }& E22 & 41 &    708/1370&    14:50:28&66&  39&   2.58&   2&  128&   yes  & yes  \\
{   }& E22 & 42 &    703/1356&    14:50:50&90&  45&   4.02&   no&  \dots&   yes  & yes  \\
{   }& E22 & 43 &    705/1367&    14:50:54&77&  110&   8.41&   no&  \dots&   yes  & yes  \\
{   }& E23 & 44 &    987/827&    14:52:14&59&  34&   1.99&   no&  \dots&   yes  & yes  \\
{   }& E23 & 45 &    987/818&    14:50:25&67&  54&   3.45&  2&  110&   yes  & yes  \\
{   }& E23 & 46 &   971/845&    14:51:05&73&  50&   3.69&   2&  68&   yes  & yes  \\
{   }& E23 & 47 &    971/835&    14:52:30&75&  52&   3.98&   no&  \dots&   yes  & yes  \\
{   }& E24 & 48 &    1439/731&    14:50:58&76&  59&   4.4&   2&  38&   yes  & yes  \\
{   }& E25 & 49 &    1392/603&    14:50:02&88&  41&   3.65&   no&  \dots&   yes  & yes  \\
{   }& E25 & 50 &    1389/603&    14:51:29&75&  42&   3.1&   no&  \dots&   yes  & yes  \\
{   }& E26 & 51 &    1396/1433&    14:51:40&50&  63&   3.13&   no&  \dots&   yes  & yes  \\
{   }& E26 & 52 &    1385/1433&    14:50:39&45&  65&   2.68&   2&  55&   yes  & yes  \\
{   }& E26 & 53 &    1401/1430&    14:51:57&50&  33&   1.63&   no&  \dots&   yes  & yes  \\
{   }& E27 & 54 &    1523/1113&    14:49:05&79&  34&   2.64&   2&  65&   yes  & yes  \\
{   }& E28 & 55 &    875/522&    14:50:34&102&  45&  4.52&   no&  \dots&   yes  & yes  \\
{   }& E29 & 56 &    583/710&    14:50:30&67&  29&   1.94&   no&  \dots&   yes  & yes  \\
{   }& E30 & 57 &   910/1301&    14:50:13&86&  39&  3.34&   no&  \dots&   yes  & yes  \\
{   }& E31 & 58 &    552/1063&    14:51:52&71&  48&   3.44&   no&  \dots&   yes  & yes  \\
{   }& E32 & 59 &    1037/1110&    14:49:20&88&  31&   2.68&   no&  \dots&   no  & yes  \\
\enddata
\tablecomments{Location: the location of a EUV upflow-like event in the pixel coordinate of EUI 174 \AA~image. Repeated: if a EUV upflow reveals a signature of recurrent. Time gap: the time interval between repeated upflow-like events.
Network: if a EUV upflow-like event originates from the network field.}
\end{deluxetable}

\begin{figure}[ht!]
\centering
\includegraphics[width=0.7\textwidth]{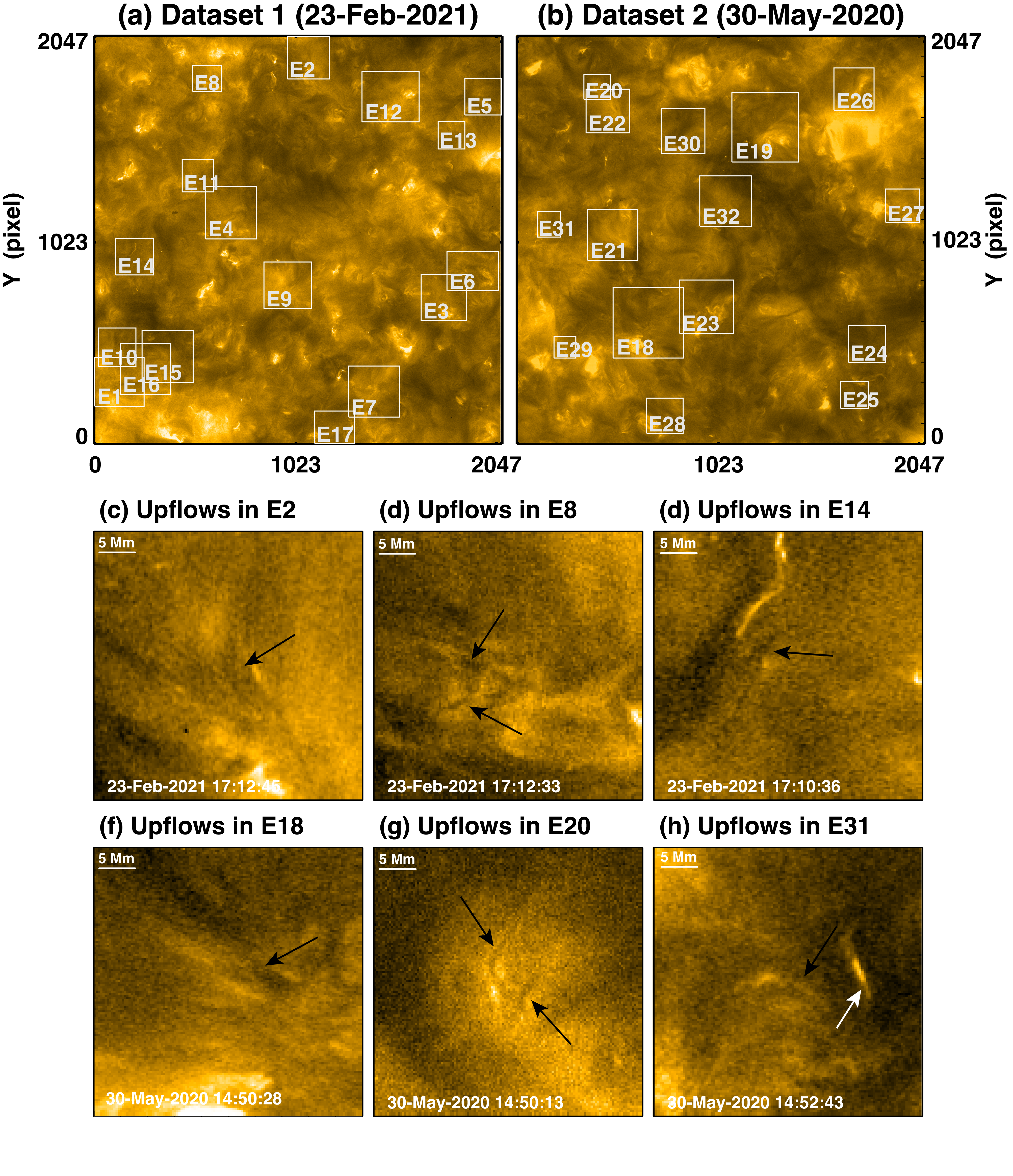}
\caption{The overview of our two datasets. In total, 32 Events FOVs are given, in which 59 upflow-like events can be detected. Six clear cases are shown in panels (c) and (d). The EUV upflow-like events of interest are marked by black arrows. For comparison, adjacent braiding of loops is marked by white arrows in panels (h).
\label{fig:general}}
\end{figure}

\begin{figure}[ht!]
\plotone{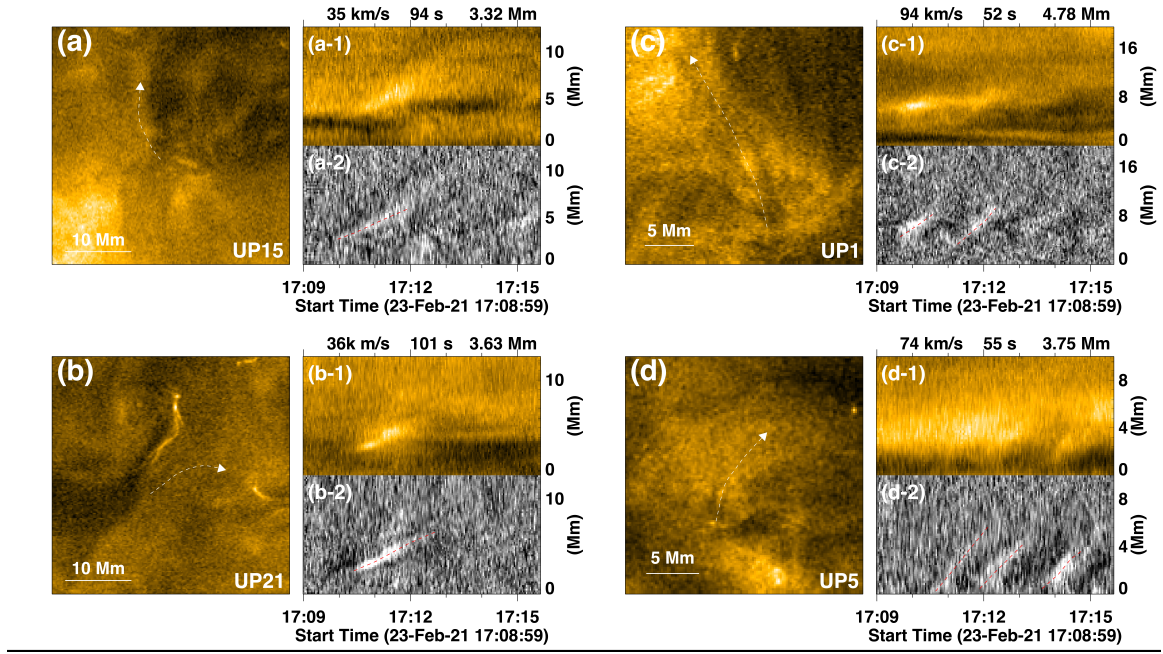}
\caption{Time-distance plot of selected EUV upflow-like events. 
(a)$-$((d)) the slit location; (a-1)$-$(d-1) original time-distance plot of EUI observation; (a-2)$-$(d-2) enhanced time-distance plot with high pass filtering.
\label{fig:general}}
\end{figure}

\begin{figure}[ht!]
\plotone{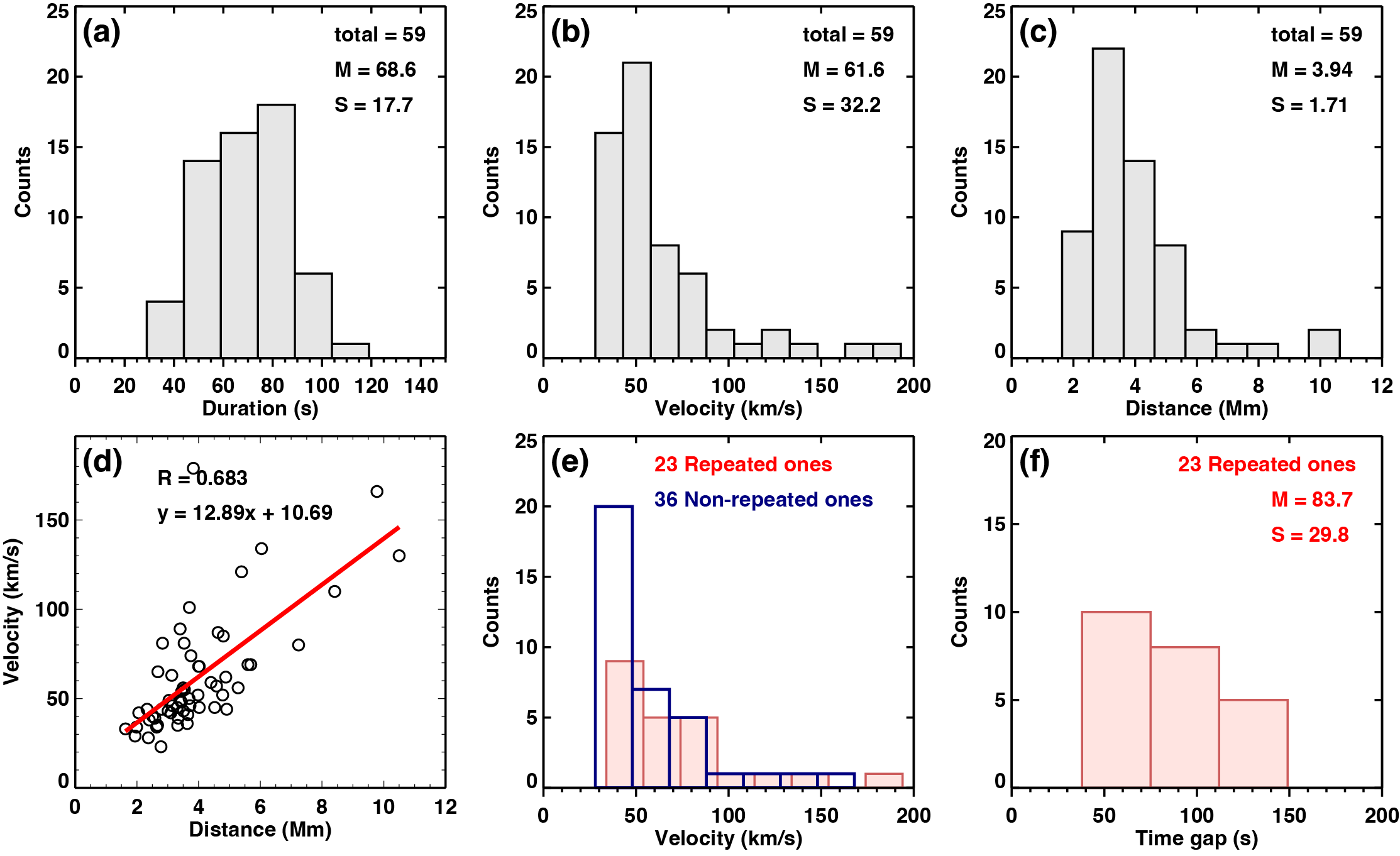}
\caption{(a-c): distributions of the parameters for EUV upflow-like events. In each panel, ‘M’and ‘S’represent the averaged value and standard deviation, respectively.  (d): the scatter plot of the relationship between the velocity  and the propagation distance for EUV upflow-like events.  (e): distributions of velocity for both repeated and non-repeated EUV upflow-like events. (f): distributions of repeated time gap for repeated upflow-like events.
\label{fig:general}}
\end{figure}

\begin{figure}[ht!]
\centering
\includegraphics[width=0.6\textwidth]{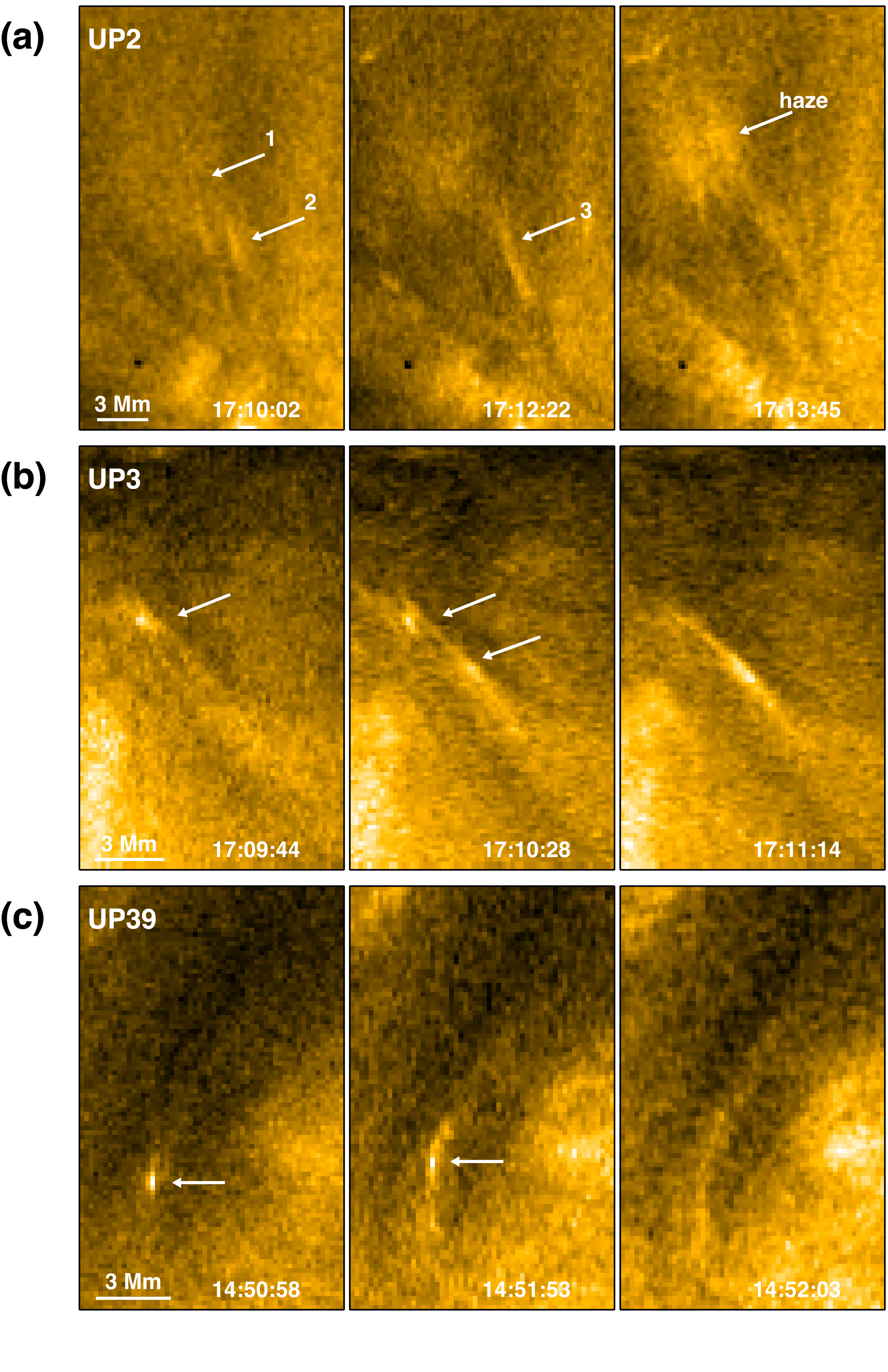}
\caption{Fine structure and evolution of EUV upflow-like events shown by three selected examples. (a): the arrows 1, 2 and 3 denote three plasma outward within a EUV upflow-like event. At 17:13:45 UT, the arrow point out a haze-like plasma. (b)$-$(c): the white arrows denote the plasma outward flows. 
\label{fig:general}}
\end{figure}

\begin{figure}[ht!]
\plotone{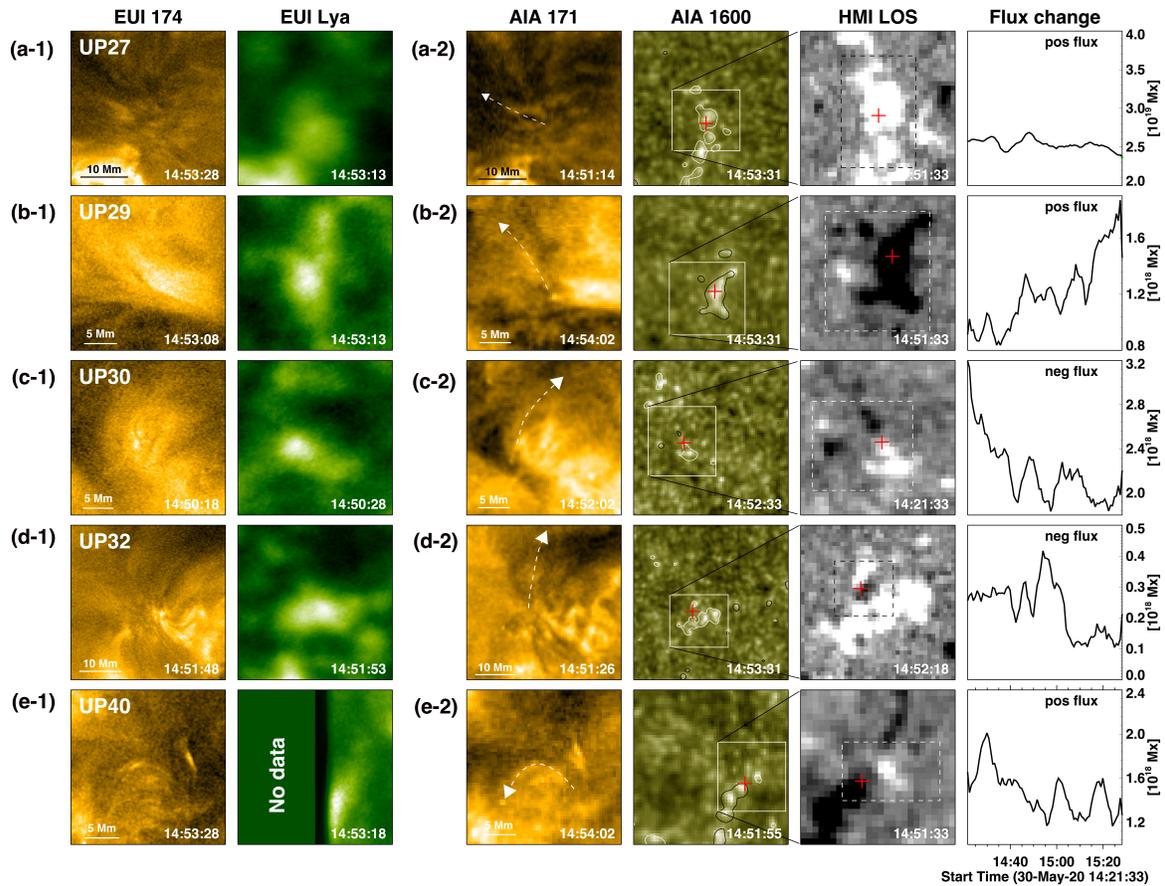}
\caption{The formation environment of EUV upflow-like events.  (a-1)$-$(e-1): EUI 174 \AA~and Ly$\alpha$ images showing five selected examples. (a-2)$-$(e-2): the corresponding five examples in SDO AIA 171\AA~, 1600\AA~images and HMI magnetograms. The white arrows denote the orientation of ejection for the EUV upflow-like events. The field of view of HMI magnetograms is indicated by the white box in AIA 1600\AA~. 
The red crosses indicate the sit where the EUV upflow-like events start. The white and black contours in AIA 1600\AA~represent magnetic field strength with the levels of ±30 G, respectively. The positive and negative flux variations in time are indicated in the black or white dashed box in HMI magnetograms.
\label{fig:general}}
\end{figure}

\begin{figure}[ht!]
\plotone{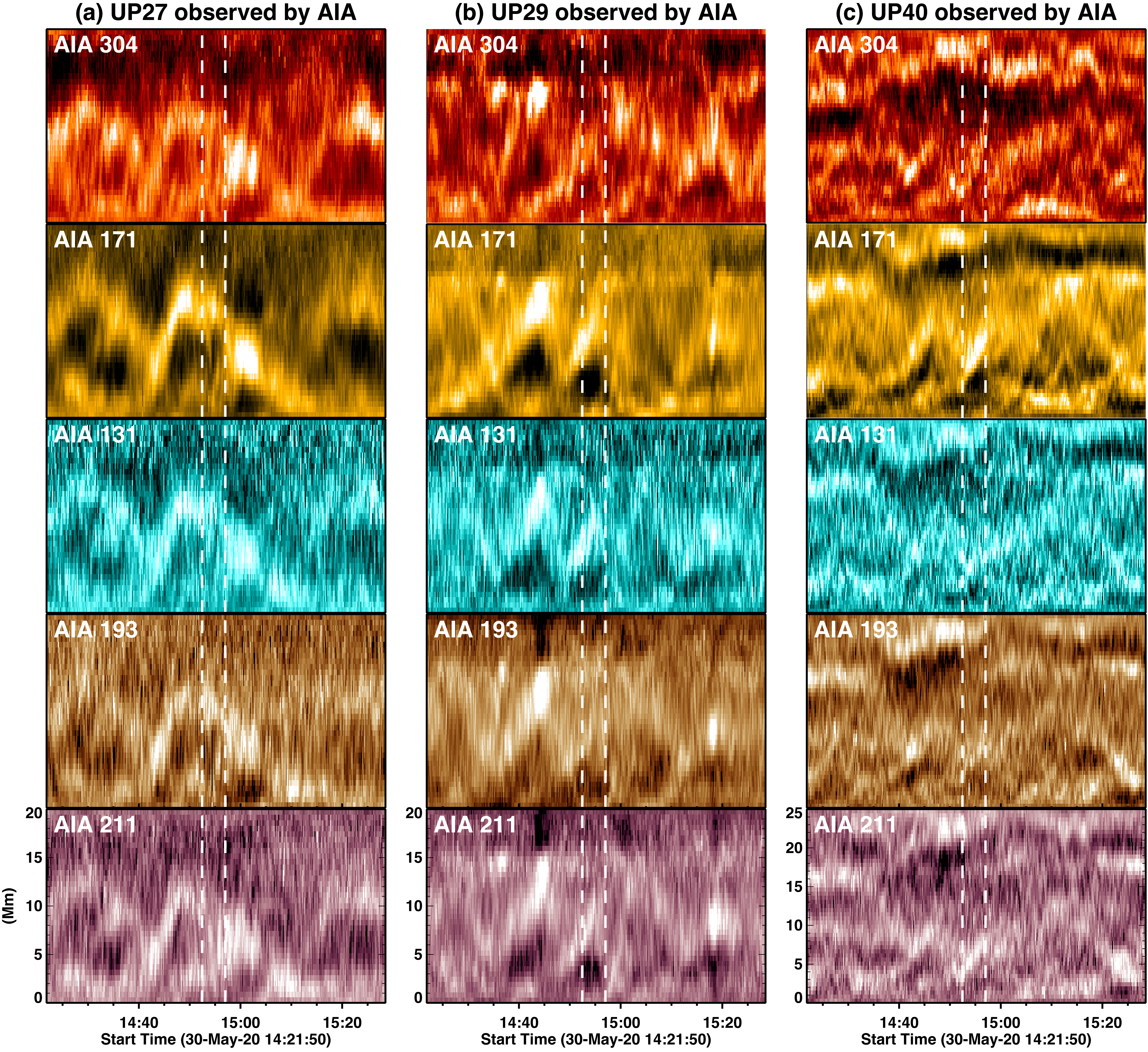}
\caption{The AIA response of three selected EUV upflow-like events. These time-distance images are plotted along the three  selected the main axis of EUV upflow-like events. The start and end time of EUI observation are marked by two dashed lines.
\label{fig:general}}
\end{figure}

\begin{figure}[ht!]
\plotone{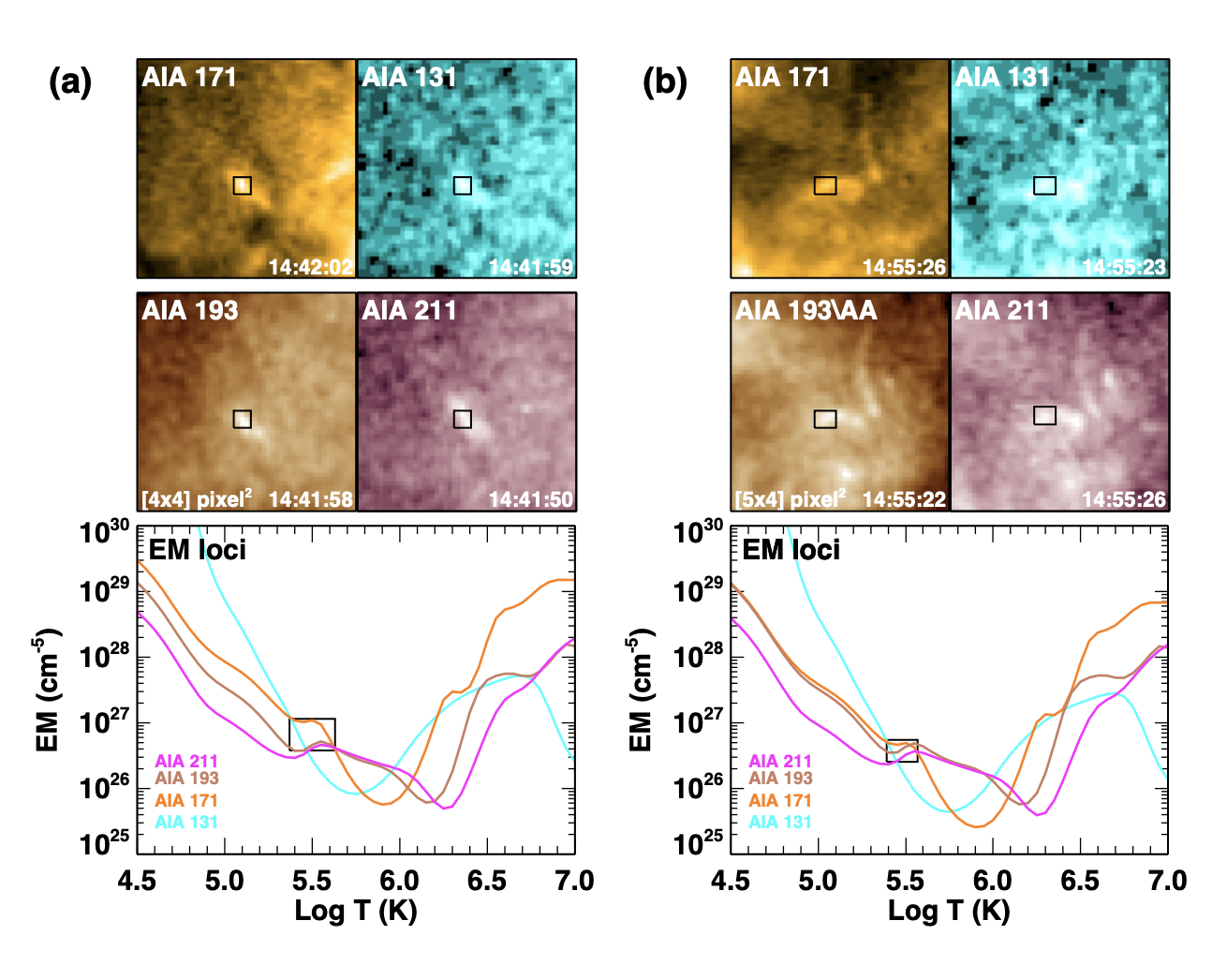}
\caption{The black boxes indicate the region for an EM-loci analysis in the bright front of the two selected upflow-like events in AIA 171, 131, 193, 211\AA~in Figure 7 (a)$-$(b). The bottom two panels display the EM-loci curves for the bright front of the EUV upflow-like events, with the small black rectangles represent the region where the EM-loci curves intersect.
\label{fig:general}}
\end{figure}

\bibliography{reference}{}
\bibliographystyle{aasjournal}

\end{document}